\documentclass[aps,twocolumn,tightenlines,nofootinbib,prl]{revtex4}
\usepackage{amsmath,amscd,amssymb}
\usepackage{color,epsfig}
\usepackage{xfrac}
\usepackage{ulem}
\usepackage{latexsym}
\usepackage{mathrsfs}
\usepackage{graphicx}
\usepackage{bbm}      
\def\beq{\begin{equation}}
\def\eeq{\end{equation}}
\def\be{\begin{eqnarray}}
\def\ee{\end{eqnarray}}
\newcommand{\dslash}{\partial \hskip -0.6em /}
\newcommand{\Dslash}{D \hskip -0.6em /}

\newcommand{\tr}{\mbox{tr}}

\begin{document}

\title{Stable charged cosmic strings}

\author{H. Weigel$^{a)}$, M. Quandt$^{b)}$, N. Graham$^{c)}$}

\affiliation{
$^{a)}$Physics Department, Stellenbosch University,
Matieland 7602, South Africa\\
$^{b)}$Institute for Theoretical Physics, T\"ubingen University
D--72076 T\"ubingen, Germany\\
$^{c)}$Department of Physics, Middlebury College
Middlebury, VT 05753, USA}

\begin{abstract}
We study the quantum stabilization of a cosmic string by a heavy fermion
doublet in a reduced version of the standard model. We show that charged strings, 
obtained by populating fermionic bound state levels, become stable if the 
electro--weak bosons are coupled to a fermion that is less than twice as 
heavy as the top quark. This result suggests that extraordinarily
large fermion masses or unrealistic couplings are not required to bind a 
cosmic string in the standard model. Numerically we find the most favorable 
string profile to be a simple ``trough'' in the Higgs vev of radius 
$\approx 10^{-18}\,\mathrm{m}$.  The vacuum remains stable in our
model, because neutral strings are not energetically favored.
\end{abstract}

\maketitle

\paragraph{Introduction}
Various field theories suggest the existence of string--like
configurations, which are the particle physics analogues of
vortices or  magnetic flux tubes in condensed matter physics. They
are called cosmic (or Z--)strings to distinguish them from the
fundamental variables in string theory and to indicate that they
can stretch over cosmic length scales.  They can have significant cosmological
effects~\cite{Copeland:2009ga} and thus may be relevant to the early universe.  
Stable strings within the standard model of particle physics would be
particularly interesting because they could be observable today.

In the standard model, string configurations
\cite{Vachaspati:1992fi,Achucarro:1999it,Nambu:1977ag}
are not topologically stable and thus can only be stabilized
dynamically.  Here we focus on the role heavy fermions can play in
this stabilization.  Since fermions can lower their energy by binding
to the string, their binding energy can overcome the classical energy
required to form the string background. However, once we
include the contribution to the energy from bound fermions, we must
also include the contribution from the distortion of the entire
fermion spectrum, i.e.~the vacuum polarization energy,
since both contributions enter at order $\hbar$.

A string configuration with a vortex structure introduces
non--trivial behavior at spatial infinity. This property
invalidates the straightforward application of standard methods to compute
the vacuum polarization energy. Recently we have shown
how to carry out such a calculation by choosing a particular
gauge~\cite{Weigel:2009wi,Weigel:2010pf}. We are thus
in a position to consistently include fermionic contributions to
the dynamical stabilization of cosmic strings.

Naculich~\cite{Naculich:1995cb} has shown that in the
limit of weak coupling, fermion fluctuations destabilize the string.
The quantum properties of $Z$--strings have been connected to
non--perturbative anomalies~\cite{Klinkhamer:2003hz}. A first attempt
at a full calculation of the quantum corrections to the $Z$--string
energy was carried out in ref.~\cite{Groves:1999ks}.
Those authors were only able to compare the energies of two
string configurations, rather than comparing a single string
configuration to the vacuum; these limitations arise from the
non--trivial behavior at spatial infinity.
The fermionic vacuum polarization energy of the Abelian
Nielson--Oleson vortex has been estimated in ref.~\cite{Bordag:2003at}
with regularization limited to the subtraction of the divergences
in the heat--kernel expansion. Quantum energies of bosonic fluctuations
in string backgrounds were calculated in ref.~\cite{Baacke:2008sq}.
Previously, we have pursued the idea of stabilizing
cosmic strings by  populating fermionic bound states in a $2+1$ dimensional
model \cite{Graham:2006qt}. Many such bound states emerge and some
configurations even induce an exact zero--mode~\cite{Naculich:1995cb}.
Nonetheless, stable configurations were only obtained for extreme values of
the model parameters. In $3+1$ dimensions, stability is more likely
because quantization of the momentum parallel to the symmetry axis yields 
an additional multiplicity of bound states.

\paragraph{Model and Ansatz}
We consider a model of the electroweak interactions in $D=3+1$ dimensions with
some technical simplifications, which we will justify {\it a posteriori}. 
We set the Weinberg angle to zero, so
that electromagnetism is decoupled from the theory and the $SU(2)$ gauge bosons
are degenerate. We also neglect QCD interactions, though we include
the  $N_C=3$ color degeneracy in computing the fermion
contribution to the string energy.  Finally, we consider a single
heavy doublet that is degenerate in mass, neglecting CKM mixing and
mass splitting within the doublet. The classical Higgs and gauge
fields are described by the Lagrangian
\begin{eqnarray}
\mathcal{L}_{\phi,W}&=&-\frac{1}{2} \tr
\left(G^{\mu\nu}G_{\mu\nu}\right) +
\frac{1}{2} \tr \left(D^{\mu}\Phi \right)^{\dag} D_{\mu}\Phi \cr
&& \hspace{1.5cm}
- \frac{\lambda}{2} \tr \left(\Phi^{\dag} \Phi - v^2 \right)^2 \,,
\end{eqnarray}
where $\Phi$ represents the Higgs doublet 
$\phi = \left(\phi_+,\phi_0\right)$
as a matrix,
$
\Phi=\begin{pmatrix}
\phi_0^* & \phi_+ \cr -\phi_+^* & \phi_0 \end{pmatrix} \,,
$
the gauge coupling constant enters via the covariant derivative
$D^\mu=\partial^\mu-igW^\mu$, and the SU(2) field strength tensor is
$G_{\mu\nu}=\partial_\mu W_\nu-\partial_\nu W_\mu
-ig\left[W_\mu,W_\nu\right]$. We then have the fermion Lagrangian
\begin{equation}
\mathcal{L}_\Psi=i\overline{\Psi}
\left(P_L \Dslash  + P_R \dslash \right) \Psi
-f\,\overline{\Psi}\left(\Phi P_R+\Phi^\dagger P_L\right)\Psi\,,
\label{gaugelag}
\end{equation}
where the Yukawa coupling $f$ controls the strength of the
Higgs--fermion interaction, which generates the fermion mass.
Our model is thus characterized by the fermion mass $m_f = fv$, the gauge
boson mass $m_W=gv/\sqrt{2}$, the Higgs mass $m_H=2v\sqrt{\lambda}$, and the
Higgs vacuum expectation value (vev) $v$.  When we introduce the
fermionic quantum corrections, we impose on--shell renormalization
conditions, in which we hold fixed $m_H$, $v$, and the
residue of the pole  in  each particle's propagator. These choices exhaust the
available counterterms,  so we have to adjust the gauge coupling $g$
to  match the physical gauge boson mass. Since we neglect boson loops,
this renormalization scheme also leaves the fermion mass unchanged.

We construct the string as a classical background field that is
translationally invariant in the $z$-direction. We work in Weyl gauge
$W^0 = 0$ and also introduce a parameter $\xi_1$ that allows us to 
include a gauge field with winding number $n$. We set $n$ to unity in 
the actual calculations. The gauge and Higgs fields are then
\begin{eqnarray}
\vec{W}&=& n\,s\,\frac{f_G(\rho)}{g\rho}\hat{\varphi}
\begin{pmatrix}
s & i c\,{\rm e}^{-in\varphi} \cr
-i c\,{\rm e}^{in\varphi} & - s
\end{pmatrix}\cr
\quad {\rm and} \quad
\Phi&=&vf_H(\rho)
\begin{pmatrix}
s\, {\rm e}^{-in\varphi} & -i c \cr
-i c & s\, {\rm e}^{in\varphi}
\end{pmatrix}\,,
\label{string}
\end{eqnarray}
where $s={\rm sin}(\xi_1)$ and $c={\rm cos}(\xi_1)$, and $(\rho,\varphi)$
are polar coordinates in the plane perpendicular to the
string axis.  This ansatz yields the classical energy per unit length
\begin{eqnarray}
E_{\rm cl}&=&2\pi \int_0^\infty d\rho \, \rho \Biggl[
\frac{2}{g^2} \left( \frac{f_G'}{\rho}  \right)^2 +
v^2 \left( f_H' \right)^2 \cr
&&\hspace{1cm}
+ \frac{v^2}{\rho^2} f_H^2 \left( 1 - f_G \right)^2
 + \lambda v^4 \left(1 - f_H^2 \right)^2 \Biggr] \, ,
\label{eq:eclslsq}
\end{eqnarray}
where primes denote derivatives with respect to $\rho$.
Variational width parameters $w_H$ and $w_G$ enter
through the respective profile functions for each field,
\begin{equation}
f_H(\rho)=1-{\rm exp}\left(\sfrac{\rho}{w_H}\right) \,\,,\,\,
f_G(\rho)=1-{\rm exp}\left(-\sfrac{\rho^2}{w_G^2}\right)\,.
\label{eqn:profile}
\end{equation}

\paragraph{Energy Considerations}
We compute the total binding energy per unit length as a
sum of three terms:
\begin{equation}
E_{\rm tot}=E_{\rm cl}+N_C(E_{\rm vac}+E_{\rm b}) \,.
\label{eq:etot}
\end{equation}

The classical energy per unit length depends on the model parameters
and the variational parameters $w_H$, $w_G$ and $\xi_1$. The two 
contributions in eq.~(\ref{eq:etot}) proportional to $N_C$ summarize 
the fermionic effects. We measure all dimensionful quantities in 
comparison to appropriate powers of $m_f$, so that $E_{\rm vac}$ and 
$E_{\rm b}$ only depend on the ansatz parameters $w_H$, $w_G$ and
$\xi_1$.\footnote{There is a weak (logarithmic) dependence on the model
parameters introduced via the on--shell renormalization conditions; it is small
for the values of the coupling constants we consider.} QCD effects only enter
via the degeneracy factor $N_C$: since the considered energy scales
are well above the QCD scale, these interactions can be
neglected due to asymptotic freedom.

The fermionic effects are computed from the single particle Dirac Hamiltonian
in the two--dimensional subspace orthogonal to the symmetry axis of the
string\footnote{We refrain from displaying this Hamiltonian, which we extract 
from eq.~(\ref{gaugelag}). For actual computations a specific gauge must be 
adopted, complicating its presentation~\cite{Weigel:2010pf,inprep}.}. The 
profiles $f_G$ and $f_H$ act as potentials in this Hamiltonian.

The vacuum polarization energy per unit length in the string
background $E_{\rm vac}$ is the computationally most
expensive part of the calculation. It is computed from the scattering
solutions to the single particle Hamiltonian using the spectral
method \cite{Graham:2009zz,Schroder:2007xk,Weigel:2009wi}.
This part involves a number of technical subtleties associated with 
the long--ranged string potential~\cite{Weigel:2010pf,inprep}.

Finally, the single particle Hamiltonian has many bound state
solutions; for $\xi_1=\frac{\pi}{2}$ there exists an exact zero
mode. By explicitly populating these bound states, we add charge to
the string.  Numerically, we compute bound state energies in the
string background by discretizing the reduced two--dimensional system
in a finite box and diagonalizing the Hamiltonian matrix numerically.
Let $\epsilon_i\le m_f$ be an eigenvalue of the two--dimensional
Dirac Hamiltonian.  Then a state has energy
$\left[\epsilon_i^2+p^2\right]^{\sfrac{1}{2}}$, where $p$ is its
conserved momentum along the symmetry axis.  To count the populated
states, we introduce a chemical potential $\mu$ such that
${\rm min}\{|\epsilon_i|\}\le\mu\le m_f$.  States with
$[\epsilon_i^2+p^2]^{\sfrac{1}{2}}<\mu$ are filled while
states with $[\epsilon_i^2+p^2]^{\sfrac{1}{2}}>\mu$ remain
empty, which gives a Fermi momentum
$P_i(\mu)= [\mu^2-\epsilon_i^2]^{\sfrac{1}{2}}$ for each  bound state.
According to the Pauli exclusion principle we can occupy each state only
once. This yields the charge density per unit length of the string
\begin{equation}
Q(\mu)=\frac{1}{\pi}\sum_{\epsilon_i\le\mu} P_i(\mu)\,,
\label{eq:charge}
\end{equation}
where the sum runs over all bound states available for a given chemical
potential.\footnote{Ambiguities in this relation due to different
boundary conditions at the end of the string show up at subleading
order in  $1/L$ where $L$ is the length of the string and can thus
be safely ignored.}

Eq.~(\ref{eq:charge}) can be inverted to give $\mu=\mu(Q)$. In  numerical 
computations we prescribe the left--hand--side of eq.~(\ref{eq:charge}) 
and increase $\mu$ from ${\rm min}\{|\epsilon_i|\}$ until the
right--hand--side matches. From this value $\mu = \mu(Q)$,
the binding energy per unit length
\begin{equation}
E_{\rm b}(Q)=\frac{1}{\pi}\sum_{\epsilon_i\le\mu}
\int_0^{\mbox{\footnotesize $P_i(\mu)$}}
\hspace{-0.2cm}dp \left[\sqrt{\epsilon_i^2+p^2}-m_f\right]
\label{eq:ebind}
\end{equation}
can be computed as a function of the prescribed charge. In this manner
the total energy becomes a function of the charge density of the string.
Filling the available states up to a common chemical potential minimizes
$E_{\rm b}$: if the towers of states built upon two different $\epsilon_i$
had different upper limits, the energy would be lowered by moving a state
from the tower with the larger limit to that with the lower one, without
changing the charge.

Our central task is to find Higgs--gauge field configurations that
yield $E_{\rm tot}<0$ for a prescribed value of the charge density, $Q$.
In doing so, we must take care that any binding we observe is not an artifact of the
Landau pole,  which eventually sends $E_{\rm vac}$ to  minus infinity as
$w_H$ and/or $w_G$ tend to zero. It arises because in our
approximation (neglecting contributions from
bosonic loops) the model is not asymptotically free.  Once we identify
a configuration and parameter set with interesting numerical results
we use a method similar to that of ref.~\cite{Hartmann:1994ai} to
ensure that the Landau pole contribution is negligible.

\paragraph{Results}
The similarity to the standard model suggests the model parameters
\begin{equation}
g=0.72\,,\,
v=177\,{\rm GeV}\,,\,
m_{\rm H}= 140\,{\rm GeV}\,,\,
f=0.99\,.
\label{eq:parameters}
\end{equation}
The Yukawa coupling estimate is obtained from the top--quark mass
$m_t=175\,{\rm GeV}$. To consider a fourth
generation with a heavy fermion doublet that couples to
the standard model bosons, we will vary the Yukawa coupling
but keep all other model parameters fixed.

\begin{figure}[tp]
\centerline{~\hspace{-0.5cm}
\includegraphics[width=7.5cm,height=4.5cm]{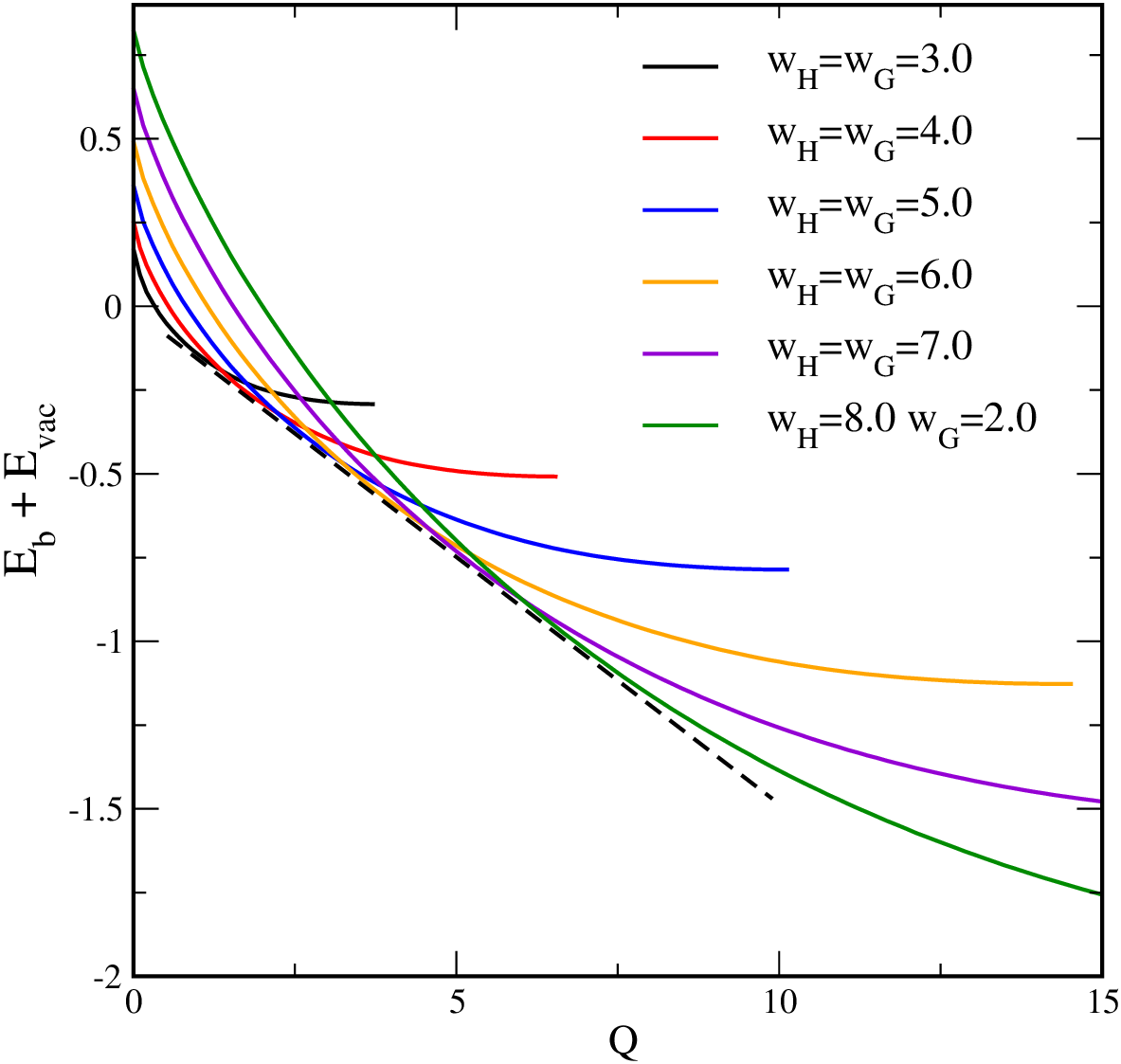}}
\caption{(Color online) Total bound state and vacuum energy per unit
length as a function of charge density per unit length, in units of
the fermion mass, for $\xi_1=0.4\pi$. The dotted line
indicates the minimal fermionic contribution to the energy.}
\label{fig:ebline}
\end{figure}

For the configurations we consider, the classical energy,
eq.~(\ref{eq:eclslsq}),
is dominated by the Higgs potential contribution, which scales as
$\lambda w_H^2/(f^4 N_C)$ compared to the fermionic
contributions.  As $\xi_1 \to 0$, the gauge field contribution goes to
zero, so this choice is favored classically.  We will see that
$\xi_1\approx 0$ remains favored when $E_{\rm vac}$ and $E_b$
are included, so that the stable charged string obtained in our model
is simply a ``trough" in the Higgs vev, without significant gauge field
contributions.

We give all numerical results in units of $m_f$ or $1/m_f$ as appropriate. 
In fig.~\ref{fig:ebline} we display the fermion contributions for various 
sets of ansatz parameters. These lines
terminate at an end point where all available bound states (for all
longitudinal momenta) are populated and the charge cannot be increased
any further. The fermion contributions favor a wide string for large charges,
while they cause the string to shrink for small charges. For very small
charges, corresponding to small widths, the calculation is unreliable
because of the Landau pole.\footnote{\label{ft_Landau} The problem arises
for widths much less than unity and coupling coupling constants of order 
five or larger. In our numerical search for stable configurations 
we only consider $w_H\ge2$ and $w_G\ge2$.}

\begin{figure}[tp]
\centerline{~\hspace{-0.5cm}
\includegraphics[width=7.5cm,height=4.5cm]{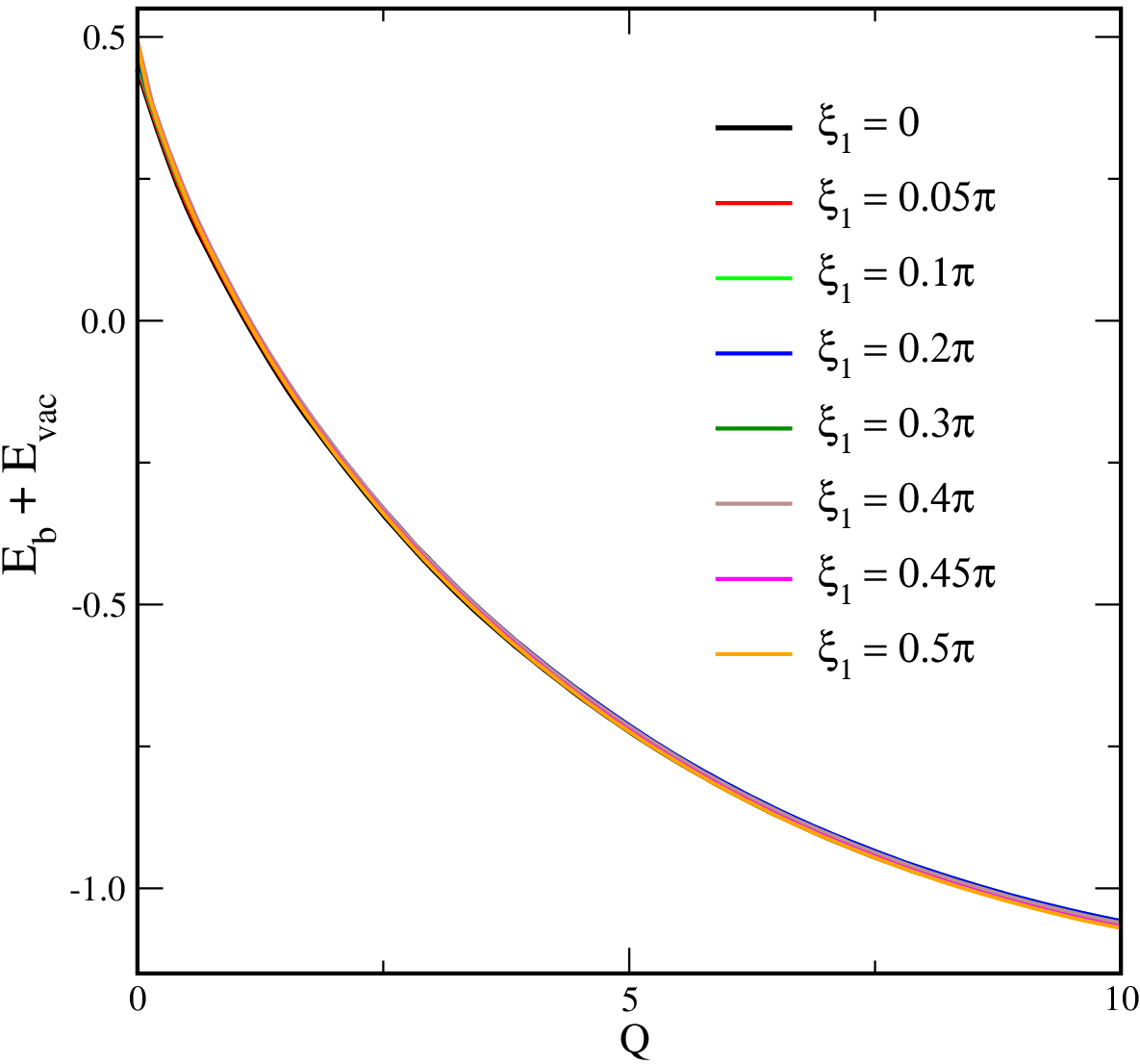}}
\caption{(Color online) Fermionic contribution to the string binding
energy per unit length as a function of charge density per unit
length, in units of the fermion mass, for a variety of values of $\xi_1$
and $w_H=6.0$ and $w_G=6.0$.}
\label{fig:xi1}
\end{figure}

When we add more configurations, we observe a linear relation between
the charge and the \emph{minimal} fermion contribution to the energy,
even though for any given  configuration, the fermion energy depends
quadratically on the charge, {\it cf.} eqs.~(\ref{eq:charge})
and~(\ref{eq:ebind}).  This linear dependence arises from a delicate
balance between the vacuum
polarization (which determines the $y$--intercept for a given
configuration) and the binding energies (which determine the $Q$
dependence).  Figure~\ref{fig:ebline} also  suggests that the width of
the Higgs profile, $w_H$, is the dominating scale
(which is corroborated in fig.~\ref{fig:xi1}, where $E_{\rm b}+E_{\rm
vac}$ is seen  to be nearly independent of $\xi_1$, and thus of
$w_G$.) Both the  number  of two--dimensional bound states
and the magnitude of their binding energies $\epsilon_i-m_f$ vary
roughly linearly with $w_H$. As a result, the minimal fermion
contribution scales quadratically with $w_H$, as the classical energy
does. To decide if the string is stable we have to compare the leading
scaling with $w_H^2$ in $E_{\rm cl}$ and $E_{\rm vac}+E_{\rm b}$. For
large widths, the string is stable if the resulting coefficient of
the scaling with $w_H^2$ is negative. For physically motivated parameters,
eq.~(\ref{eq:parameters}), the classical energy dominates and there is no
binding for any charge. However, as mentioned above, the relative contribution
from $E_{\rm cl}$ decreases like $1/f^4$. So even a moderate increase of
the fermion mass could lead to binding. 
We remark that extrapolating the straight line in
fig.~\ref{fig:ebline} predicts that the vacuum energy should vanish for 
very narrow strings, as we would expect. This estimate overcomes the Landau 
pole obstacles that arise in a direct computation.

To search for a stable string of fixed charge $Q$, we have computed the vacuum
polarization energy and the bound state energies from the
two--dimensional Hamiltonian for several hundred configurations
characterized by specific values of the ansatz parameters 
$w_H$, $w_G$ and $\xi_1$.  We then prescribe the charge $Q$ and, 
for those configurations that can accommodate it according to
eq.~(\ref{eq:charge}), we compute the binding
energy as in eq.~(\ref{eq:ebind}). Once we have computed the fermionic
contribution to $E_{\rm tot}$, the classical energy is a simple
spatial integral, which requires a negligible amount of
additional computation. As a result, in this procedure
it is most efficient to vary the Yukawa coupling. For a
given charge, we then have a large set of configurations that
are labeled by given (discrete) values of the
variational parameters. We scan this set for the minimal total
energy. If the variational parameters covered the full configuration
space, this treatment would be equivalent to the self--consistent
construction of the minimal energy configuration.  With our
restriction to the variational space, however, we only find an
upper limit to the exact minimum; if our treatment detects a 
bound configuration, the existence of a stable charged cosmic
string is established.

\begin{figure}[tp]
\centerline{~\hspace{-0.5cm}
\includegraphics[width=7.5cm,height=4.5cm]{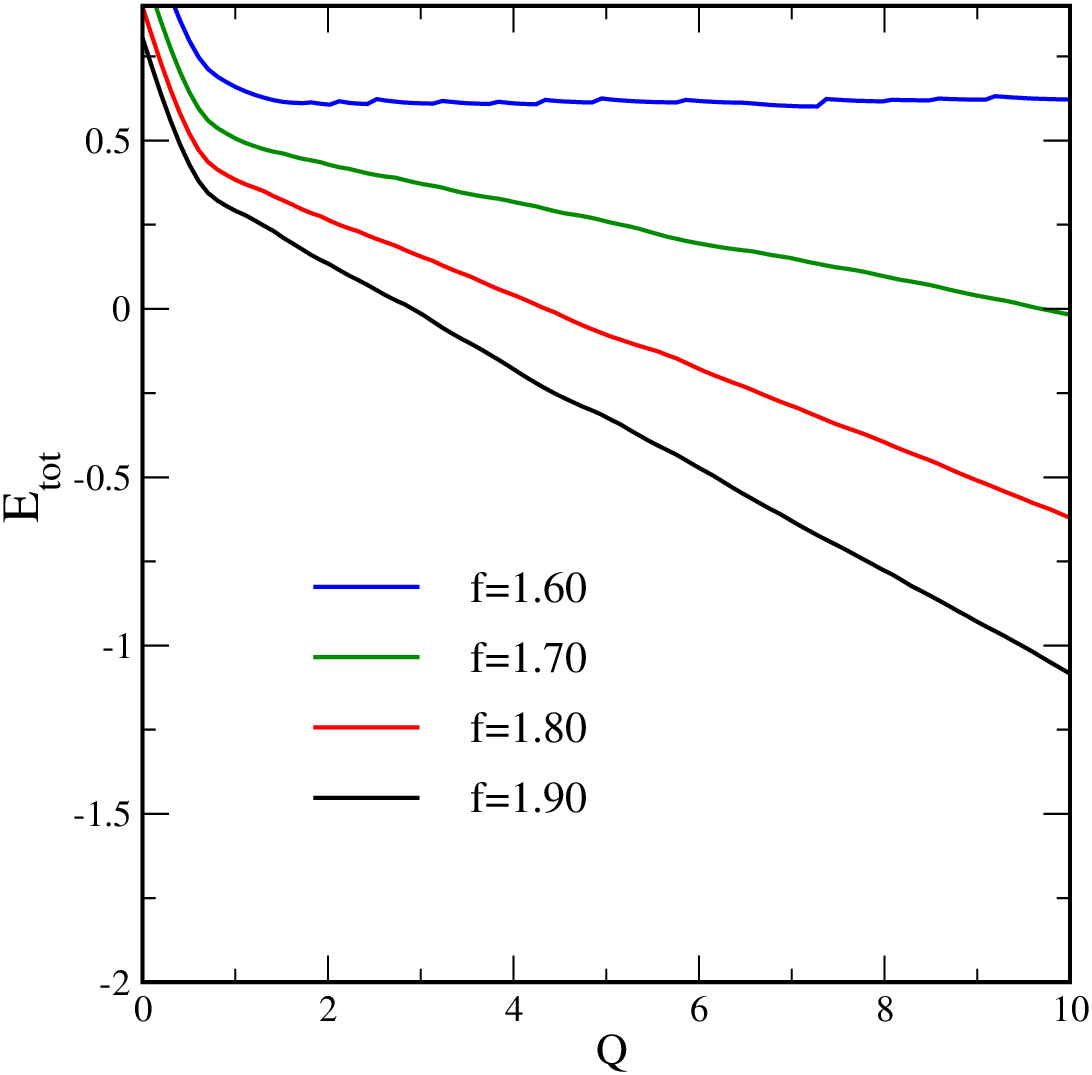}}
\caption{(Color online) Total energy per unit length of optimal string
configurations as a function of charge per unit length, in
units of the fermion mass.}
\label{fig:etot1}
\end{figure}

In fig.~\ref{fig:etot1}, we show the full energy per unit length
$E_{\rm tot}$ as a function of the charge density per unit length for
a variety of Yukawa couplings $f$. The sharp increase at small
$Q$ is an artifact of the restriction of the ansatz parameters, 
{\it cf.} footnote~\ref{ft_Landau}. Increasing the Yukawa coupling
from its top--quark value decreases the relative contribution from
$E_{\rm cl}$ to $E_{\rm tot}$. We see that at $f\approx 1.6$ the large
width pieces from $E_{\rm cl}$ and $E_{\rm vac}+E_{\rm b}$ approximately
cancel.  Increasing the Yukawa coupling only slightly more, \emph{e.g.}~to
$f\gtrsim 1.7$, yields a negative total energy per
unit length at large charge densities, which indicates that the string is
lighter than the corresponding density of free fermions. This limit
corresponds to a fermion mass of about $300\,{\rm GeV}$ with a typical
width for the stable charged string of about $10^{-18}{\rm m}$
($w_H\approx 4/m_f$).

Surprisingly, we find that the fermion contribution to the energy is 
nearly independent of the ansatz parameter $\xi_1$, as shown in
fig.~\ref{fig:xi1}. Even though the bound state spectrum varies strongly 
with $\xi_1$, and $E_{\rm vac}+E_{\rm b}$ depends only weakly on
$\xi_1$\cite{Weigel:2010pf}, there are subtle cancellations
within the bound state spectrum itself that yield such a tiny
gauge field dependence of the  fermion energy. As $g\ll f$, the gauge
field terms increase $E_{\rm cl}$ for $\xi_1 \neq 0$. Hence $E_{\rm tot}$
is minimized for $\xi_1\approx0$ in the cases we have studied.

\paragraph{Discussion}
We have seen that a heavy fermion doublet can stabilize a nontrivial string
background in a simplified version of the electroweak standard model for
a non--zero fixed charge density. Light fermions would contribute only weakly 
to the binding of the string, since their Yukawa couplings are small. As a result, 
we can add them to our model, {\it e.g.} to accommodate anomaly cancellation, 
without significantly changing the result. The resulting configuration is essentially 
of pure Higgs structure. Any additional (variational) degree of freedom can only 
lower the total energy. Hence embedding this configuration in the full standard model, 
with the $U(1)$ gauge field included, also yields a bound object. We see 
binding set in at $m_f \approx 300\,\mathrm{GeV}$, which is still within the range 
of energy scales at which the standard model should provide an effective description 
of the relevant physics, and also within the range to be probed at the LHC. For
such fermion masses, recent calculations have also suggested
the potential stability of multi--fermion bound states in a Higgs
background~\cite{Froggatt:2008hc,Kuchiev:2010ux}.  

The fermion bound states carry non--zero angular momenta, implying
that the bound state wave--functions depend on the azimuthal angle. This
might induce a more complicated spatial structure of the string configuration
than the one adopted in eq.~(\ref{string}). In particular, the cylindrical analog 
of spherical ``hedgehog'' configurations, representing a Higgs field with unit 
winding within a U(1) subgroup of the full SU(2) isospin group, could be an 
interesting extension of our work.  Such alterations can only lower the 
total energy, however.

\paragraph{Acknowledgments}
N.~G.\ is supported in part by the NSF through grant PHY08-55426.

\end{document}